\documentclass[aps,prb,twocolumn,showpacs]{revtex4}
\usepackage{color}
\usepackage{graphicx}
\usepackage{ulem}

\begin{document}

\title{Superconductivity in electron-doped arsenene}
\author{Xin Kong$^{1}$}
\author{Miao Gao$^{2}$}
\email{gaomiao@nbu.edu.cn}
\author{Xun-Wang Yan$^{3}$}
\author{Zhong-Yi Lu$^{4}$}
\author{Tao Xiang$^{1,5}$}
\date{\today}
\affiliation{$^{1}$Institute of Physics, Chinese Academy of Sciences, Beijing 100190,
China }
\affiliation{$^{2}$Department of Microelectronics Science and Engineering, Faculty of
Science, Ningbo University, Zhejiang 315211, China}
\affiliation{$^{3}$College of Physics and Engineering, Qufu Normal University, Shandong
273165, China}
\affiliation{$^{4}$Department of Physics, Renmin University of China, Beijing 100872,
China}
\affiliation{$^{5}$Collaborative Innovation Center of Quantum Matter, Beijing, China}

\begin{abstract}
  Based on the first-principles density functional theory electronic structure calculation, we investigate the possible phonon-mediated superconductivity in arsenene, a two-dimensional buckled arsenic atomic sheet, under electron doping.
   We find that the strong superconducting pairing interaction results mainly from the $p_z$-like electrons of arsenic atoms and the $A_1$ phonon mode around the $K$ point, and the superconducting transition temperature can be as high as 30.8 K in the arsenene with 0.2 doped electrons per unit cell and 12\% applied biaxial tensile strain.
  This transition temperature is about ten times higher than that in the bulk arsenic under high pressure.
  It is also the highest transition temperature that is predicted for electron-doped two-dimensional elemental superconductors, including graphene, silicene, phosphorene, and borophene.
\end{abstract}

\maketitle

\section{INTRODUCTION}

Recently there has been a surge of interest in the investigation of
two-dimensional (2D) superconductors, partially due to their potential
application in nano-superconducting devices~\cite%
{Franceschi-Nature5,Huefner-PRB79}.
A pure 2D electronic system can be obtained by growing a single layer graphite or other materials on a proper substrate.
The interplay between a 2D superconductor and the substrate has proven to be an efficient way
to enhance superconductivity.
For example, the superconducting transtion temperature, T$_c$, of a single layer  FeSe film deposited on the (001) surface of SrTiO$_3$ is greatly enhanced~\cite{Wang-CPL29,Ge-Nat14} in comparison with the bulk FeSe superconductor~\cite{Hsu-PNAS105}, resulting from the coupling between electrons
and the phonon modes in the substrate~\cite{Lee-Nature515}.

Superconductivity in graphene, which is the first 2D compound synthesized at
laboratory, was extensively explored. Through a plasmon-mediated mechanism,
Uchoa~\textit{et al.} discussed properties of superconducting states in
several metal coated graphenes~\cite{Uchoa-PRL98}. First-principles density functional theory (DFT) calculations predicted that the monolayer LiC$_{6}$ and CaC$_{6}$ are
phonon-mediated superconductors with T$_{c}$ of 8.1 K and 1.4 K,
respectively~\cite{Profeta-Nature8}. Later, the superconducting phase was
observed experimentally below 7.4 K in a Li-intercalated graphite thin film
\cite{Tiwari-arXiv} and 6 K in a Ca-decorated graphene \cite{Chapman-arXiv}.
Calculations of electron-phonon interactions suggested that T$_{c}$ of
graphene can reach 23.8 K or even 31.6 K upon heavy electron or hole doping
under 16.5\% biaxial tensile strain (BTS) \cite{Si-PRL111}, but experimental
evidence for such high-T$_{c}$ superconductivity in doped graphenes is still not
available.

In recent years, several new 2D materials, like silicene, phosphorene, and
borophene, were synthesized experimentally. Similar to graphene, phosphorene
is obtained by mechanically exfoliating layered black phosphorus \cite%
{Liu-Nano8}. Silicene \cite{Lalmi-APL97,Chen-PRL109} and borophene \cite%
{Mannix-Science,Feng-arXiv1} were epitaxially grown on Ag(111) surfaces. It
was reported that there is a charge transfer from the Ag(111) substrate to
silicene \cite{Chen-PRL110} or borophene \cite{Feng-arXiv1}. Furthermore,
the substrate imposes strain to these single-layer materials, due to the
lattice mismatch \cite{Feng-arXiv1,Nie-APL94,Ding-Nano10,Conley-Nano13}.
These effects of the substrate should be taken into account in the
investigation of superconducting properties in these materials.

The DFT calculation also showed that the superconducting transition temperature can reach 16.4 K in silicene upon electron doping of $n_{\text{2D}}$ = 3.51$%
\times $10$^{14}$ cm$^{-2}$ and 5\% BTS \cite{Wan-EPL104}. T$_{c}$ of
phosphorene was predicted to be 12.2 K under the doping of 2.6$\times $10$%
^{14}$ cm$^{-2}$ and 8\% uniaxial tensile strain along the armchair
direction \cite{Shao-EPL108}. Applying 4\% BTS to phosphorene can further
increase T$_{c}$ to 16 K \cite{Ge-NJP17}.

Unlike silicene and phosphorene,
borophene is intrinsically a metal \cite{Feng-arXiv1}. A free-standing
pristine borophene is predicted to superconduct around 20 K \cite%
{Penev-Nano16,Gao-PRB95}, especially for the $\chi _{3}$-type borophene
whose T$_{c}$ can be 24.7 K \cite{Gao-PRB95}. This is the highest
superconducting transition temperature among predicted or observed 2D
elemental superconductors without doping. Unfortunately, the charge transfer
and tensile strain imposed by the Ag substrate suppress the superconducting
order dramatically \cite{Gao-PRB95,Cheng-2D4}.

Recently, a buckled single-layer honeycomb arsenic, i.e. arsenene, was
proposed \cite{Zhang-Angew54,Kamal-PRB91}. Unlike the semi-metallic bulk
gray arsenic, arsenene is a semiconductor with an indirect energy gap of
2.49 eV. More importantly, arsenene undergoes an intriguing
indirect-to-direct gap transition by applying a small BTS. This makes arsenene
a promising candidate of transistor with high on/off ratios, optoelectronic
device working under blue or UV light, and 2D-crystal-based mechanical
sensor \cite{Zhang-Angew54}. Furthermore, it was predicted that arsenene has
a higher electron mobility than that of MoS$_{2}$ \cite{Pizzi-Nat7}, and can
become a unique topological insulator under suitable BTS without considering
any spin-orbit coupling \cite{Zhang-Nanoscale7}.

In this work, we employ the first-principles DFT and the Wannier
interpolation technique to accurately determine the electron-phonon coupling (EPC) properties of electron-doped arsenenes. The
phonon-mediated superconducting T$_{c}$ is evaluated based on the
McMillian-Allen-Dynes formula. Without applying BTS, 0.1
electrons/cell (hereafter $e$/cell) doping can already turn arsenene into a superconductor with a superconducting temperature above the liquid-helium temperature. T$_{c}$ increases
to 10 K at 0.3 $e$/cell doping (i.e. $n_{\text{2D}}$=2.76$%
\times $10$^{14}$ cm$^{-2}$). Moreover, we find that the $A_{1}$ phonon mode at the $K$
point contributes mostly to the EPC constant. From the electronic point of
view, the $p_{z}$-like electronic orbitals of arsenic atoms couple strongly with
phonons. By applying a BTS, the $A_{1}$ phonon mode is softened and the EPC matrix
elements are enhanced. This enlarges both the EPC $\lambda $ and the superconducting transition temperature. Under a 0.2 $e$/cell doping and
12\% BTS, T$_{c}$ of arsenene is predicted to be about 30.8 K. To the
best of our knowledge, this is the highest T$_{c}$ predicted for 2D elemental
superconductors upon electron doping.

\section{COMPUTATIONAL APPROACH}

In the DFT-based electronic structure calculation, the plane wave basis method is adopted \cite{pwscf}. We
calculate the Bloch states and the phonon perturbation potentials \cite%
{Giustino-PRB76} using the local density approximation and the
norm-conserving pseudopotentials \cite{Troullier-PRB43}. The kinetic energy
cut-off and the charge density cut-off are taken to be 80 Ry and 320 Ry,
respectively. A slab model is used to simulate arsenene, in which a 12 {\AA }
vacuum is added to avoid the interaction between the neighboring arsenic sheets
along the $c$-axis. Electron doping is simulated by adding electrons into
the system with a compensating background of uniform positive charges. For
each doping concentration, the atomic positions are relaxed but with fixed
in-plane lattice constants which are obtained by optimizing the lattice
structure of arsenene without doping.

The charge density is calculated on an unshifted mesh of 60$\times $60$\times $1 points, with
a Methfessel-Paxton smearing \cite{Methfessel-PRB40} of 0.02 Ry. The dynamical matrix and perturbation potential
are calculated on a $\Gamma $-centered 12$\times $12$\times $1 mesh, within
the framework of density-functional perturbation theory \cite%
{Baroni-RMP73_515}. Maximally localized Wannier functions \cite%
{Marzari-PRB56,Souza-PRB65} are constructed on a 12$\times $12$\times $1
grid of the Brillouin zone, using ten random Gaussian functions as the
initial guess. Fine electron (600$\times $600$\times $1) and phonon (200$%
\times $200$\times $1) grids are used to interpolate the EPC constant using
the Wannier90 and EPW codes \cite{Mostofi-CPC178,Noffsinger-CPC181}. Dirac $%
\delta $-functions for electrons and phonons are replaced by smearing
functions with widths of 15 and 0.2 meV, respectively.

The EPC constant $\lambda $ is determined by the
summation of the momentum-dependent coupling constant $\lambda _{\mathbf{q}%
\nu }$ over the first Brillouin zone or the integration of the Eliashberg
spectral function $\alpha ^{2}F(\omega )$ in the frequency space \cite%
{Allen-PRB6_2577,Allen-RPB12_905},
\begin{equation}
\lambda =\frac{1}{N_{q}}\sum_{\mathbf{q}\nu }\lambda _{\mathbf{q}\nu }=2\int
\frac{\alpha ^{2}F(\omega )}{\omega }d\omega,  \label{eq:lambda}
\end{equation}%
in which $N_{q}$ represents the total number of \textbf{q} points in the fine
\textbf{q}-mesh. The coupling constant $\lambda_{\mathbf{q}\nu }$ for mode $\nu $ at
wavevector $\mathbf{q}$ is defined by \cite{Allen-PRB6_2577,Allen-RPB12_905}%
,
\begin{equation}
\lambda _{\mathbf{q}\nu }=\frac{2}{\hbar N(0)N_{k}}\sum_{ij\mathbf{k}}\frac{1%
}{\omega _{\mathbf{q}\nu }}|g_{\mathbf{k},\mathbf{q}\nu }^{ij}|^{2}\delta
(\epsilon _{\mathbf{q}}^{i})\delta (\epsilon _{\mathbf{k+q}}^{j}).
\label{eq:lambda_qv}
\end{equation}%
Here $\omega _{\mathbf{q}\nu }$ is the phonon frequency and $g_{\mathbf{k},%
\mathbf{q}\nu }^{ij}$ is the probability amplitude for scattering an
electron with a transfer of crystal momentum $\mathbf{q}$. $\left(
i,j\right) $ and $\nu $ denote the indices of energy bands and phonon modes,
respectively. $\epsilon _{\mathbf{q}}^{i}$ and $\epsilon _{\mathbf{k+q}}^{j}$
are the eigenvalues of the Kohn-Sham orbitals with respect to the Fermi
energy. $N(0)$ is the density of states (DOS) of electrons at the Fermi
level. $N_{k}$ is the total number of \textbf{k} points in the fine
Brillouin-zone mesh. The Eliashberg spectral function is determined by \cite%
{Allen-PRB6_2577,Allen-RPB12_905},
\begin{equation}
\alpha ^{2}F(\omega )=\frac{1}{2}\sum_{\mathbf{q}\nu }\delta (\omega -\omega
_{\mathbf{q}\nu })\lambda _{\mathbf{q}\nu }\omega _{\mathbf{q}\nu }.
\label{eq:spectral}
\end{equation}

We calculate the superconducting transition temperature using the
McMillian-Allen-Dynes formula \cite{Allen-RPB12_905},
\begin{equation}
T_{c}=f_{1}f_{2}\frac{\omega _{\text{log}}}{1.2}\exp \left[ \frac{%
-1.04(1+\lambda )}{\lambda (1-0.62\mu ^{\ast })-\mu ^{\ast }}\right].
\label{eq:Tc}
\end{equation}%
$f_{1}$ and $f_{2}$ are the correction factors, which equal $1$ when $%
\lambda <1.3$, and
\begin{eqnarray}
f_{1} &=&[1+(\lambda /\Lambda _{1})^{3/2}]^{1/3}, \\
f_{2} &=&1+\frac{(\sqrt{\langle \omega ^{2}\rangle }/\omega _{\text{log}%
}-1)\lambda ^{2}}{\lambda ^{2}+\Lambda _{2}^{2}}
\end{eqnarray}%
when $\lambda \geq 1.3$. Here $\Lambda _{1}$=$2.46(1+3.8\mu ^{\ast })$, $%
\Lambda _{2}$=$1.82(1+6.3\mu ^{\ast })(\sqrt{\langle \omega ^{2}\rangle }%
/\omega _{\text{log}})$. The effective screened Coulomb repulsion constant $%
\mu ^{\ast }$ is set to 0.1, and
\begin{eqnarray*}
\omega _{\text{log}} &=&\exp \left[ \frac{2}{\lambda }\int \frac{d\omega }{%
\omega }\alpha ^{2}F(\omega )\log \omega \right] , \\
\langle \omega ^{2}\rangle &=&\frac{2}{\lambda }\int d\omega \alpha
^{2}F(\omega )\omega .
\end{eqnarray*}

\section{RESULTS AND DISCUSSION}

\subsection{EPC in electron doped arsenene}

\begin{figure}[t]
\begin{center}
\includegraphics[width=8.6cm]{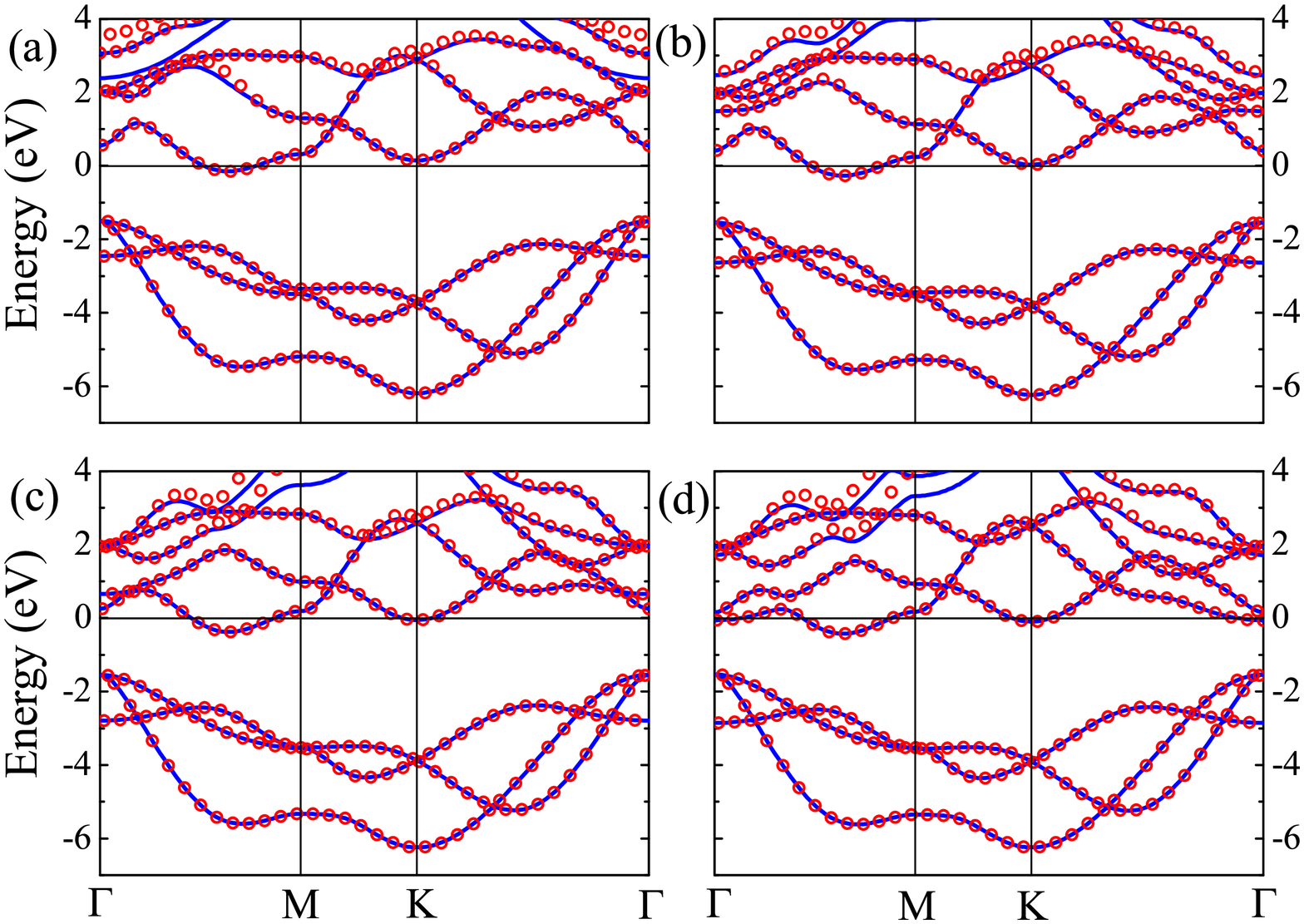} %
\includegraphics[width=8.6cm]{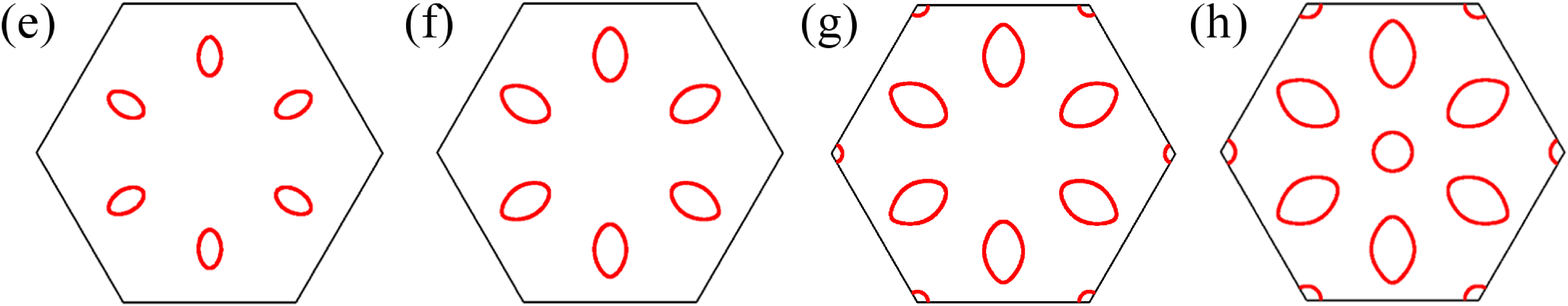}
\end{center}
  \caption{(Color online) Band structures of electron-doped arsenenes at four doping levels: (a) 0.1 $e$/cell, (b) 0.2 $e$/cell,  (c) 0.3 $e$/cell, and (d) 0.4 $e$/cell. The Fermi set to zero. The solid blue lines represent the band structures obtained by first-principles DFT calculations. The red circles denote the band structures obtained by interpolation with MLWFs. (e-h) the Fermi surfaces correspond to the band structures shown in (a-d), respectively.
  }
\label{fig:band-e}
\end{figure}

For a free standing arsenene without doping, we find that the optimized lattice
constant is 3.5411 {\r{A}}, in agreement with the result, 3.5408 {\r{A}},
obtained by Zhang \textit{et al.} \cite{Zhang-Angew54}. Our calculation shows
that arsenene is semiconducting with an indirect band gap of 1.42 eV. The
valence band maximum is at the $\Gamma$ point, while the conduction band
minimum is located on the line between $\Gamma$ and $M$. Although the
energy gap is underestimated in comparison with the HSE06-level result \cite{Zhang-Angew54}, it does not affect our EPC results since the shapes of
conduction band given by LDA and HSE06 are concordant.

Figure~\ref{fig:band-e} shows the band structures with the corresponding Fermi surfaces for four electron-doped
arsenenes. At the doping of 0.1 $e$/cell, there are six elliptical electron
pockets surrounding $\Gamma $. With the increase of the doping level, the
area enclosed by each pocket expands. Moreover, six small electron arcs
emerge at the Fermi level around the six zone corners when the doping level reaches 0.3 $e$/cell or above [Fig.~\ref{fig:band-e}(g)]. At higher doping, an electron pocket centred at the $\Gamma $ point appears at the
Fermi level [Fig.~\ref{fig:band-e}(h)]. The influence of electron doping on
fully occupied energy bands is negligible.

\begin{figure}[t]
\begin{center}
\includegraphics[width=8.6cm]{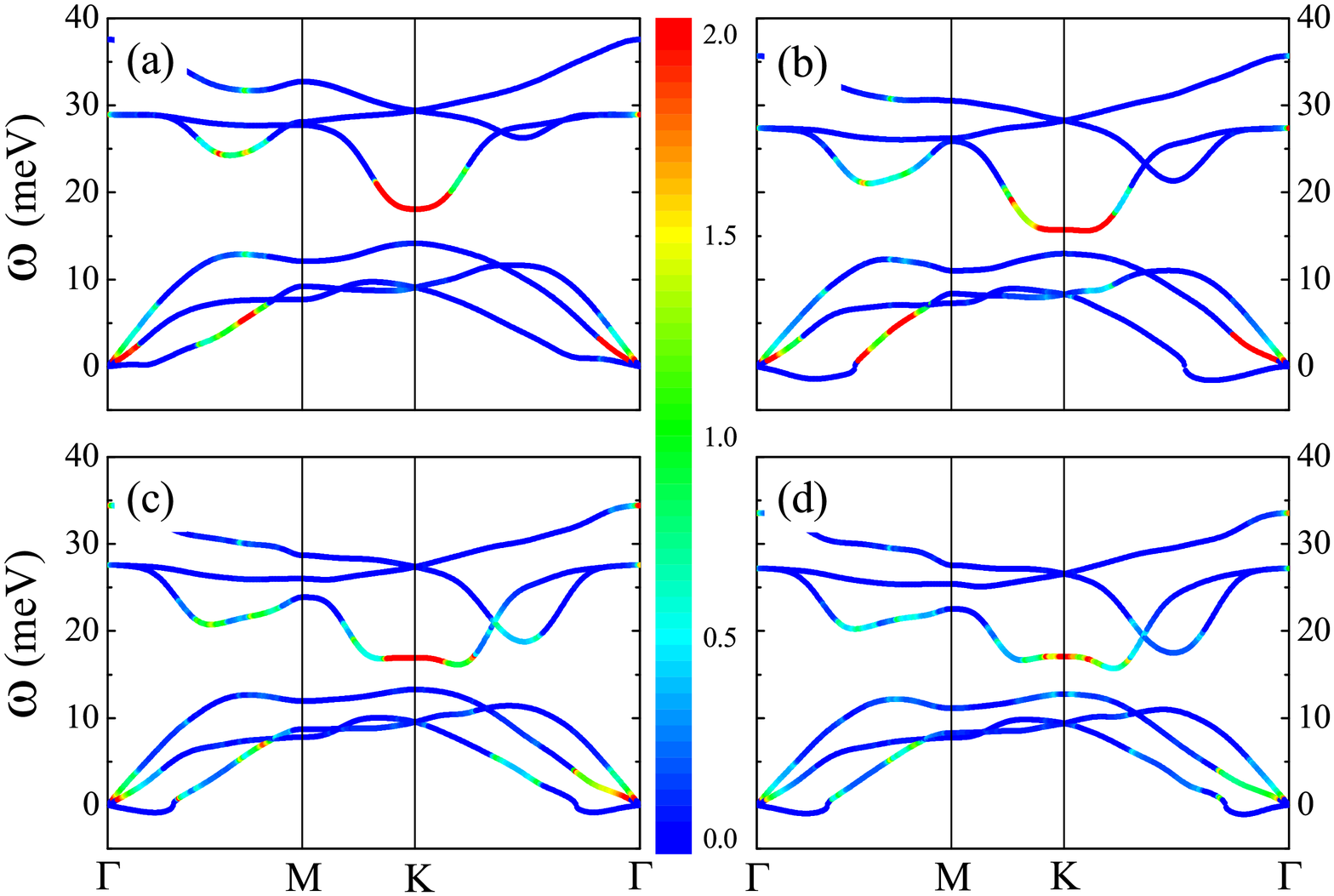}
\includegraphics[width=8.6cm]{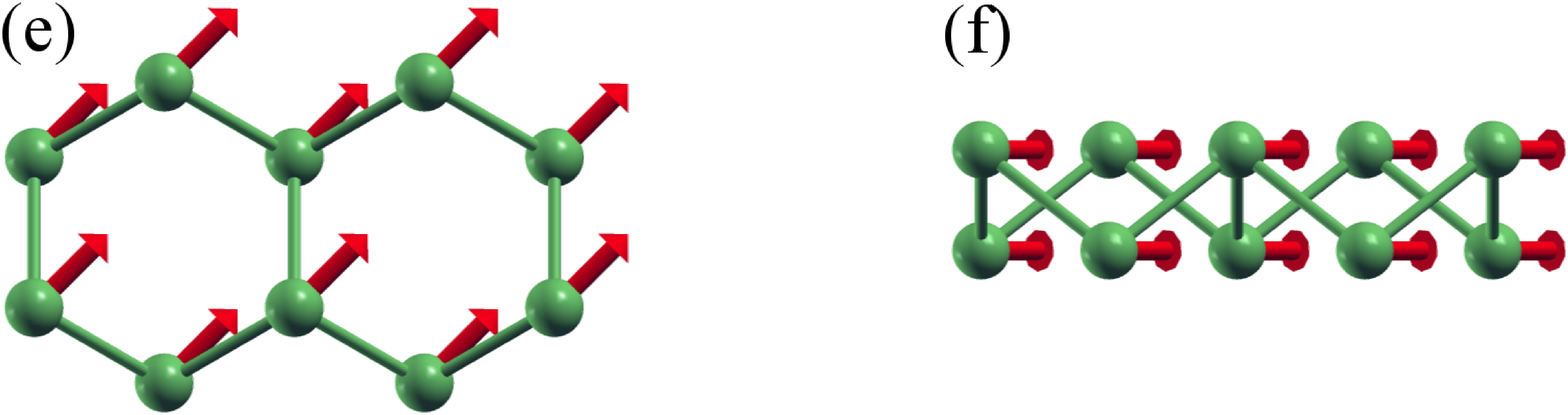}
\end{center}
\caption{
  (Color online) Phonon spectra of electron-doped arsenenes at four doping levels: (a) 0.1 $e$/cell. (b) 0.2 $e$/cell. (c) 0.3 $e$/cell. (d) 0.4 $e$/cell.
  The amplitude of the EPC constant, i.e. $\protect\lambda_{\mathbf{q\protect\nu}}$ is mapped by different colors.  (e) and (f) show the top and side views of the $A_1$ optical phonon mode in real space.}
\label{fig:phonon-e}
\end{figure}

Figure~\ref{fig:phonon-e} shows the phonon spectra of electron-doped
arsenenes. At the doping of 0.1 $e$/cell, there is no imaginary phonon
frequency, indicating that this system is dynamically stable.
Moreover, there is a gap between the acoustic and optical phonon modes.
With the increase of doping, the optical phonon modes are gradually softened.
At higher doping, imaginary frequencies are found in the lowest acoustic band near the  $\Gamma $ point
[Fig.~\ref{fig:phonon-e}(b)-\ref{fig:phonon-e}(d)].
This kind of imaginary frequencies were also found in the phonon spectra of borophene \cite{Gao-PRB95}, arsenene \cite{Kamal-PRB91,Zhang-Nanoscale7}, germanene \cite{Cahangirov-PRL102}, and other binary monolayer honeycomb materials \cite%
{Sahin-PRB80}.
It is not a sign of structure instability.
Instead, it results from the numerical instability in the accurate calculation of rapidly decreasing interatomic forces \cite{Sahin-PRB80}.
The largest contribution to $\lambda $ comes from the $A_{1}$ mode in the
lowest optical phonon mode around the $K$ point.
A real-space picture of the eigen-vector of this mode is shown in Fig.~\ref{fig:phonon-e}(e) and Fig.~\ref{fig:phonon-e}(f).

The Eliashberg spectral function $\alpha^2F(\omega)$ shows two main peaks [Fig.~\ref%
{fig:a2f-e}].
The lower-frequency peak results from the $A_1$ phonon mode around the $K$ point.
In comparison with the peak at the 0.1 $e$/cell doping, the peak position shifts towards lower frequency at higher doping.
The higher-frequency peak is also mainly the contribution of the lowest optical phonon excitations, especially those between $\Gamma$ and $M$.
Even though the EPC $\lambda_{\mathbf{q}\nu}$ of these phonons along $\Gamma$-$M$ is smaller than that of the $A_1$ mode at the $K$ point, the density of states is higher.

\begin{figure}[t]
\begin{center}
\includegraphics[width=8.6cm]{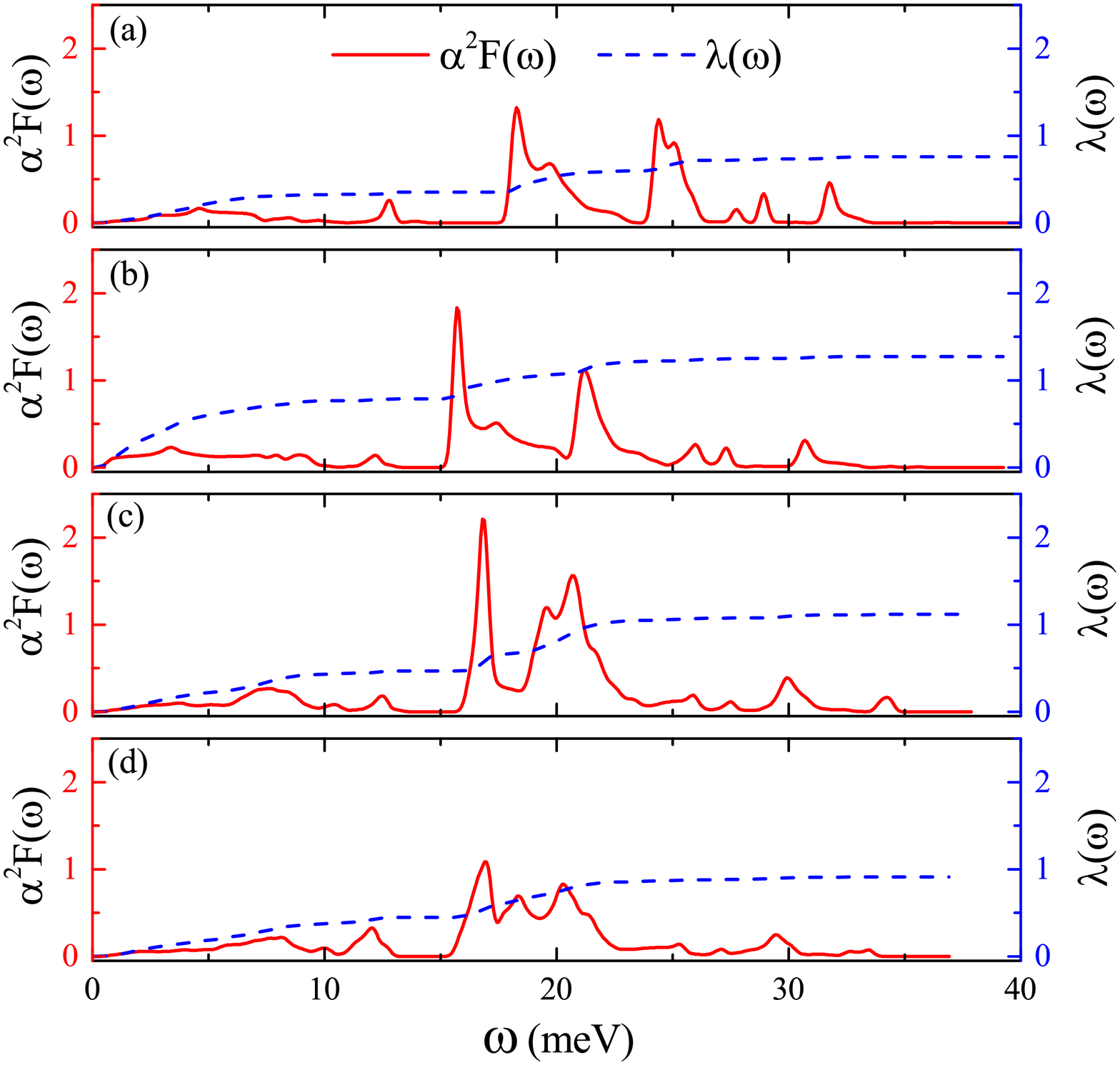}
\end{center}
  \caption{(Color online) Frequency dependence of the Eliashberg spectral function $\alpha^2F(\omega)$ and $\lambda(\omega)$ for four electron-doped arsenenes: (a) 0.1 $e$/cell. (b) 0.2 $e$/cell. (c) 0.3 $e$/cell. (d) 0.4 $e$/cell. $\lambda(\omega)$ is defined by $2\int_0^\omega \frac{1}{\omega'}\alpha^2F(\omega')d\omega'$. }
\label{fig:a2f-e}
\end{figure}

\begin{table}[bp]
\caption{
  Predicted superconducting transition temperature T$_c$ and other parameters obtained by the first-principles calculation at various doping  ($e$/cell) and BTS $\varepsilon$ (\%): $N$(0) (states/spin/eV/cell), frequency of $A_1$ phonon $\omega (A_1)$ (meV), $\lambda$, $\omega_{\text{log}}$ (meV), $\sqrt{\langle\omega^2\rangle}$ (meV), and T$_c$ (K). }
\label{table:energy}%
\begin{tabular}{cccccccc}
\hline\hline
doping & $\varepsilon$ & $N$(0) & $\omega(A_1)$ & $\lambda$ & $\omega_{\text{%
log}}$ & $\sqrt{\langle\omega^2\rangle}$ & T$_c$ \\ \cline{1-8}
0.1 & 0.0 & 0.37 & 18.08 & 0.76 & 9.81 & 16.85 & 4.7 \\
0.2 & 0.0 & 0.40 & 15.72 & 1.27 & 5.65 & 12.85 & 6.3 \\
0.3 & 0.0 & 0.61 & 16.93 & 1.12 & 10.66 & 16.21 & 10.1 \\
0.4 & 0.0 & 0.91 & 17.08 & 0.91 & 9.86 & 15.16 & 6.9 \\
0.2 & 2.0 & 0.45 & 14.89 & 1.45 & 6.46 & 12.94 & 9.7 \\
0.2 & 4.0 & 0.54 & 11.18 & 1.90 & 6.57 & 11.85 & 13.6 \\
0.2 & 6.0 & 0.64 & 7.83 & 2.77 & 6.88 & 10.40 & 19.6 \\
0.2 & 8.0 & 0.77 & 6.06 & 4.27 & 5.99 & 8.86 & 24.4 \\
0.2 & 12.0 & 1.16 & 7.85 & 6.12 & 5.41 & 8.53 & 30.8 \\ \hline\hline
\end{tabular}%
\end{table}

Based on the above results, we calculate the superconducting transition temperature T$_c$ using the McMillian-Allen-Dynes formula \cite{Allen-RPB12_905}.
The results, together with other key parameters, are presented in Table I.
Without BTS, T$_c$ shows a maximum at 0.3 $e$/cell. $\lambda$ at 0.2 $e$/cell is larger than that at 0.3 $e$/cell due to the considerable contribution from the acoustic phonon modes at the former doping [Fig.~\ref{fig:phonon-e}(b)], but the corresponding T$_c$ is lower than the latter case.

The EPC constant $\lambda$ is determined by the DOS, the phonon frequency $\omega_{\mathbf{q}\nu}$, the EPC matrix element $|g_{\mathbf{k}, \mathbf{q} \nu}^{ij}|$, and other parameters. In order to determine which effect has the largest contribution to $\lambda$, we calculate the following two quantities
\begin{eqnarray}
\xi(\mathbf{q}) &=& \frac{1}{N(0)N_k}\sum_{ij\mathbf{k}}\delta(\epsilon_{\mathbf{%
k}}^i)\delta(\epsilon_{\mathbf{k+q}}^j),
\\
\gamma(\mathbf{q}) & = & \frac{1}{N(0)N_k}\sum_{ij\nu\mathbf{k}} |g_{\mathbf{k},%
\mathbf{q}\nu}^{ij}|^2\delta(\epsilon_{\mathbf{q}}^i)\delta(\epsilon_{%
\mathbf{k+q}}^j).
\end{eqnarray}
$\xi(\mathbf{q})$ is a modified Fermi surface nesting function, in which $N(0)$ is also
included.
$\gamma(\mathbf{q})$ is the nesting function weighted by the EPC matrix element $|g_{\mathbf{k},\mathbf{q}\nu}^{ij}|$.

\begin{figure}[t]
\begin{center}
\includegraphics[width=8.6cm]{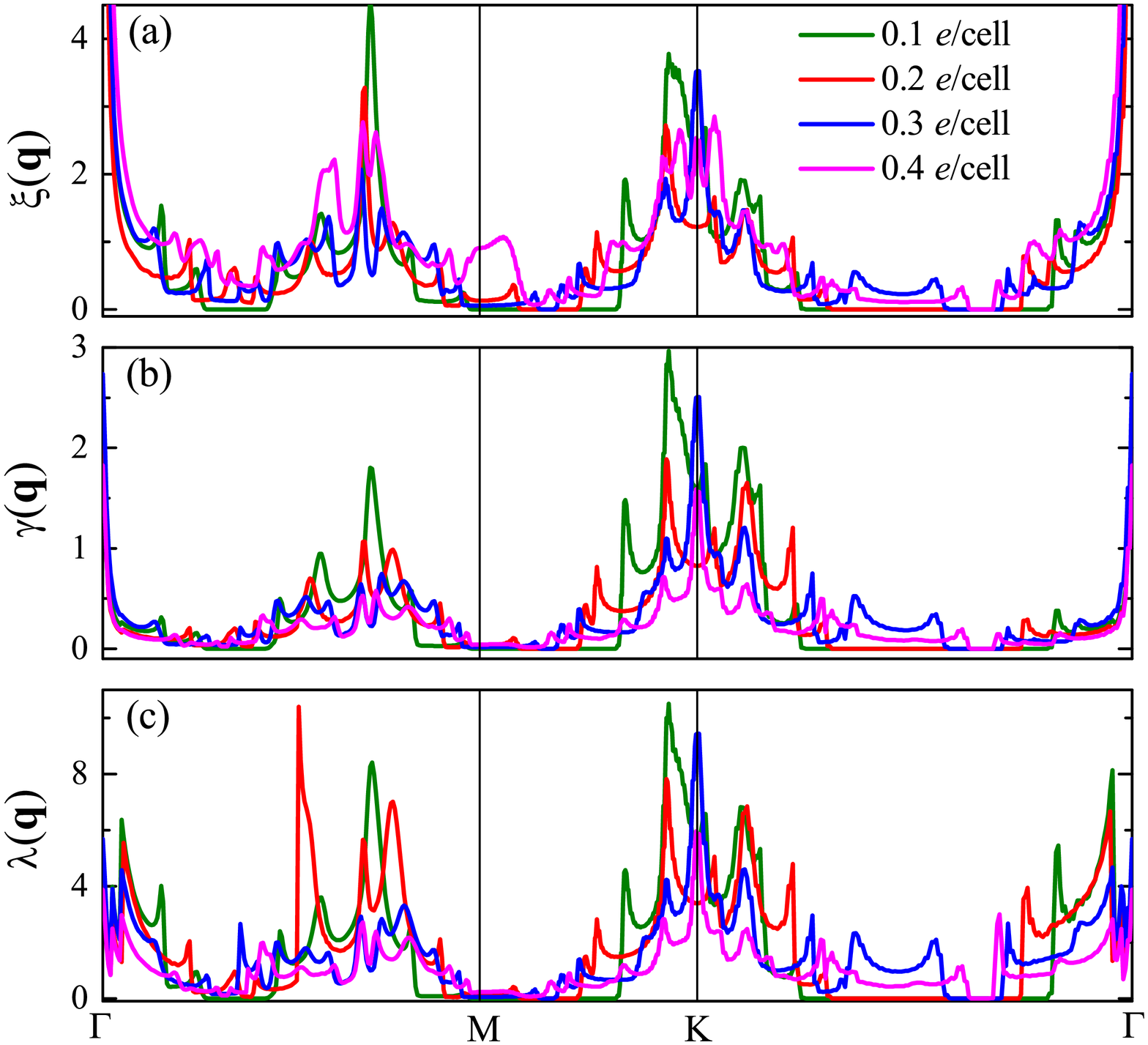} %
\includegraphics[width=8.6cm]{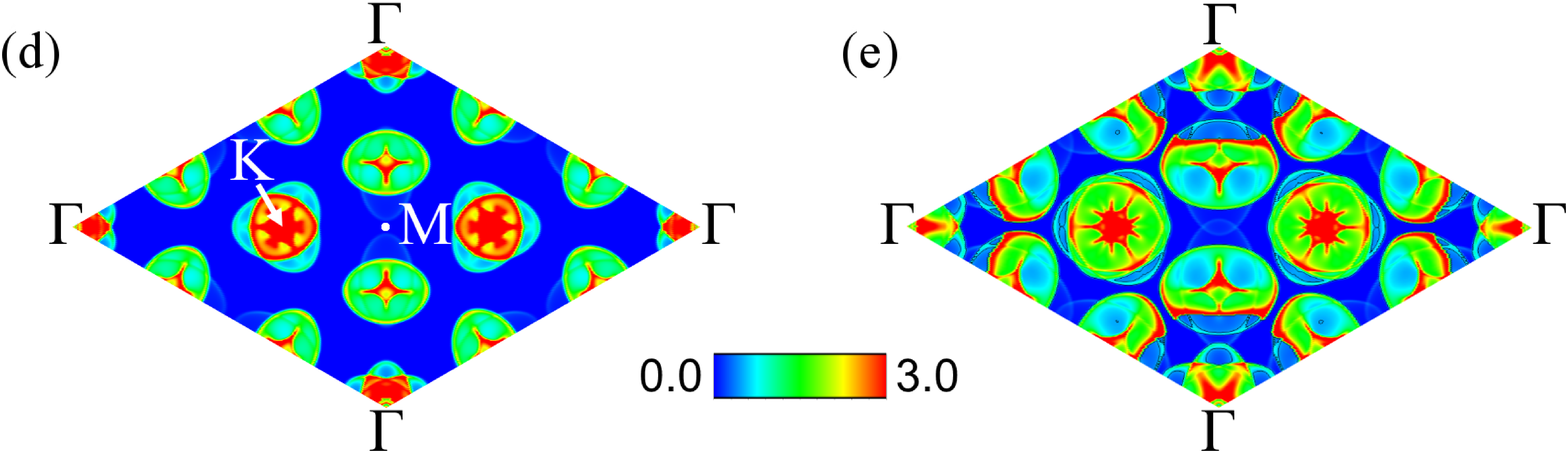}
\end{center}
\caption{(Color online) Modified Fermi surface nesting function $\protect\xi(%
\mathbf{q})$ (a), $\protect\gamma(\mathbf{q})$ (b), and $\protect\lambda(%
\mathbf{q})$ (c) of electron-doped arsenenes. (d) and (e) represent the $%
\protect\lambda(\mathbf{q})$ in the whole Brillouin zone for 0.1 $e$/cell
and 0.2 $e$/cell doping, respectively. It is noted that the reciprocal unit
cell instead of the first Brillouin zone is used.}
\label{fig:nesting-e}
\end{figure}

Figure \ref{fig:nesting-e} shows $\xi(\mathbf{q})$, $\gamma(\mathbf{q})$,
and $\lambda(\mathbf{q})$ (calculated through $\sum_{\nu}\lambda_{\mathbf{%
q\nu}}$) for the four electron-doped arsenenes.
At each doping level, the similarity between $\xi(\mathbf{q})$ and $\lambda(\mathbf{q})$ indicates that the strong EPC of the $A_1$ phonon mode mainly comes from the peak in the
nesting function $\xi(\mathbf{q})$ around the $K$ point.
The relatively lower vibrational frequencies of strongly coupled phonon modes at the 0.2 $e$/cell doping lead to the sharp peaks between $\Gamma$ and $M$ and the largest $\lambda $.
$\lambda(\mathbf{q})$ at 0.1 $e$/cell is comparable to that at 0.2 $e$/cell [Fig.~\ref{fig:nesting-e}(c)]. However, $\lambda$ is a summation over the whole Brillouin zone. For the cases of 0.2, 0.3, and 0.4 $e$/cell doping, there is a certain amount of $\mathbf{q}$ points which have substantial contribution to $\lambda(\mathbf{q})$ [Fig.~\ref{fig:nesting-e}(d) and Fig.~\ref{fig:nesting-e}(e)] but are not located along the $\Gamma$-$M$-$K$-$\Gamma$ high-symmetry line.

\subsection{EPC under BTS}

\begin{figure}[t]
\begin{center}
\includegraphics[width=8.6cm]{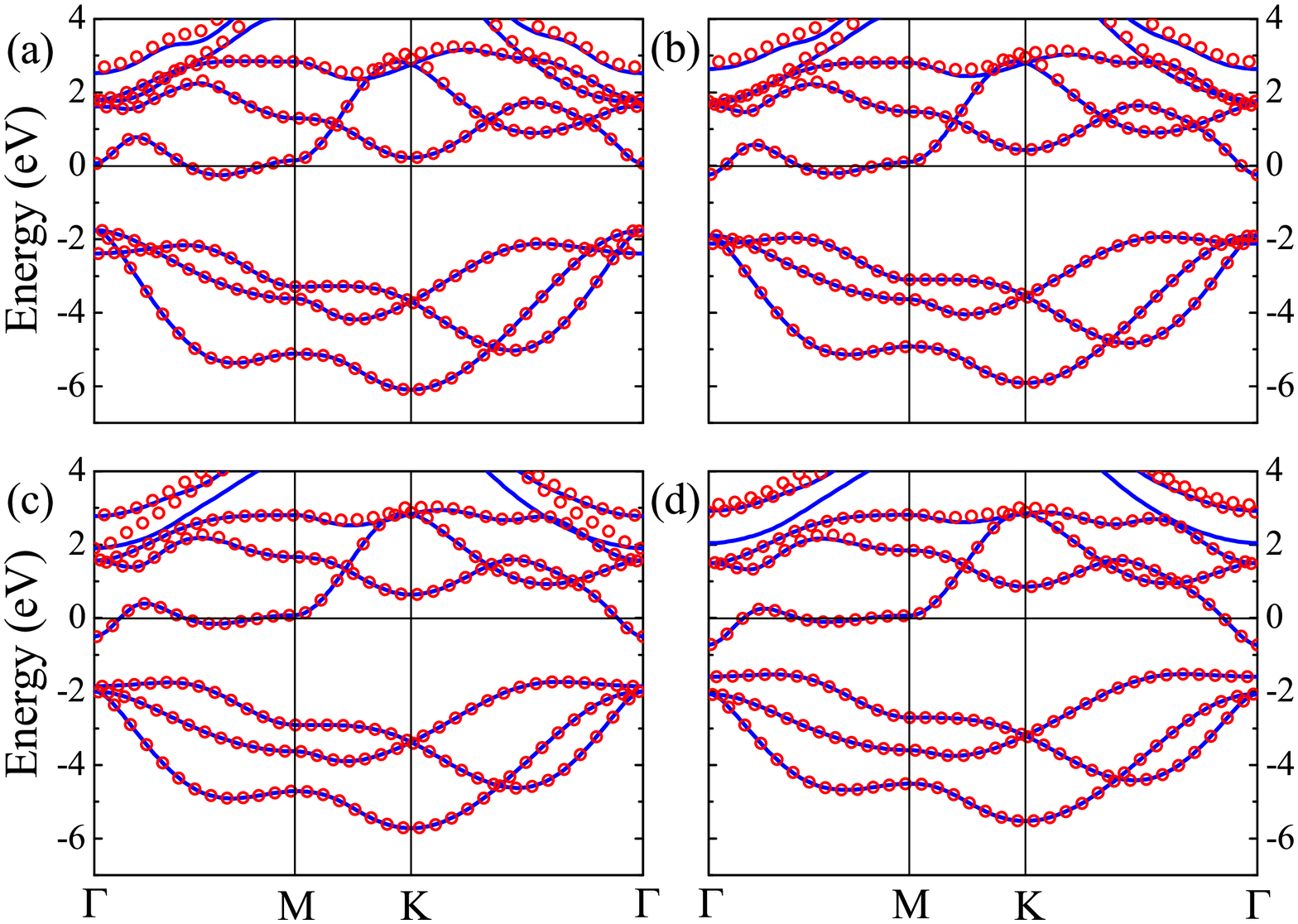} %
\includegraphics[width=8.6cm]{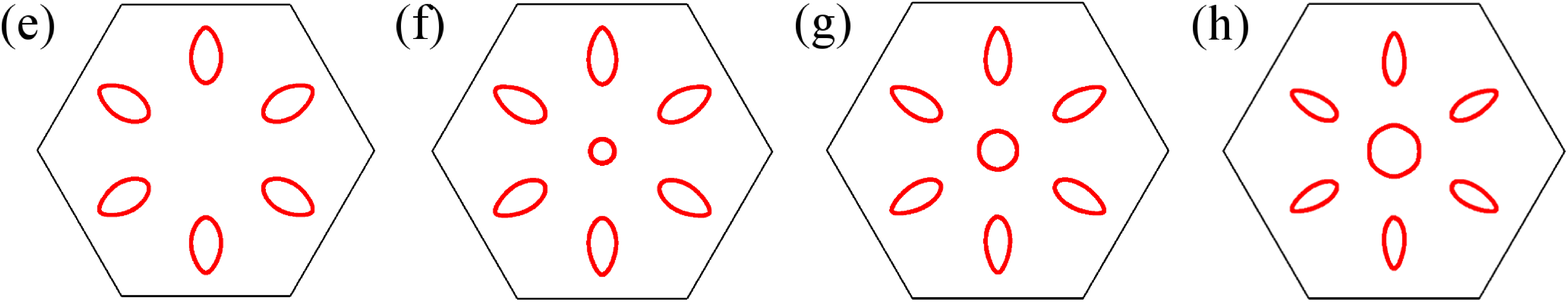}
\end{center}
  \caption{(Color online) Band structures of arsenenes with 0.2 $e$/cell electron doping under (a) 2\%, (b) 4\%, (c) 6\%, and (d) 8\% BTS. (e-h) the Fermi surfaces correspond to the band structures shown in (a-d), respectively.  }
\label{fig:band-s}
\end{figure}

The BTS is measured by $\varepsilon=(a-a_0)/a_0\times100\%$, where $a_0$ and $a$ are the lattice constants without and with strain, respectively.
Without doping, arsenene remains in the semiconducting phase under BTS up to 12\% \cite{Zhang-Angew54}.
In order to study the BTS effect on the superconducting properties, we calculate the EPC of biaxial strained arsenenes under 0.2 $e$/cell doping.

At 4\% BTS, as shown in Fig.~\ref{fig:band-s}(b), the conduction band minimum moves to the $\Gamma$ point.
This leads to a direct band gap of 1.67 eV, resembling the indirect-to-direct band-gap transition in the undoped arsenene under 4\% BTS \cite{Zhang-Angew54}.
With the increase of BTS, the conduction band along the $\Gamma$-$M$ line becomes less dispersive [Fig.~\ref{fig:band-s}(c-d)], which enlarges the DOS at the Fermi level [see Table I]. Furthermore, the applied BTS lowers the conducton-band energy at the $\Gamma$ point, which enables an electron pocket around the $\Gamma$ point to emerge at the Fermi surface and reduces the volume of the six elliptical Fermi surface sheets.

\begin{figure}[t]
\begin{center}
\includegraphics[width=8.6cm]{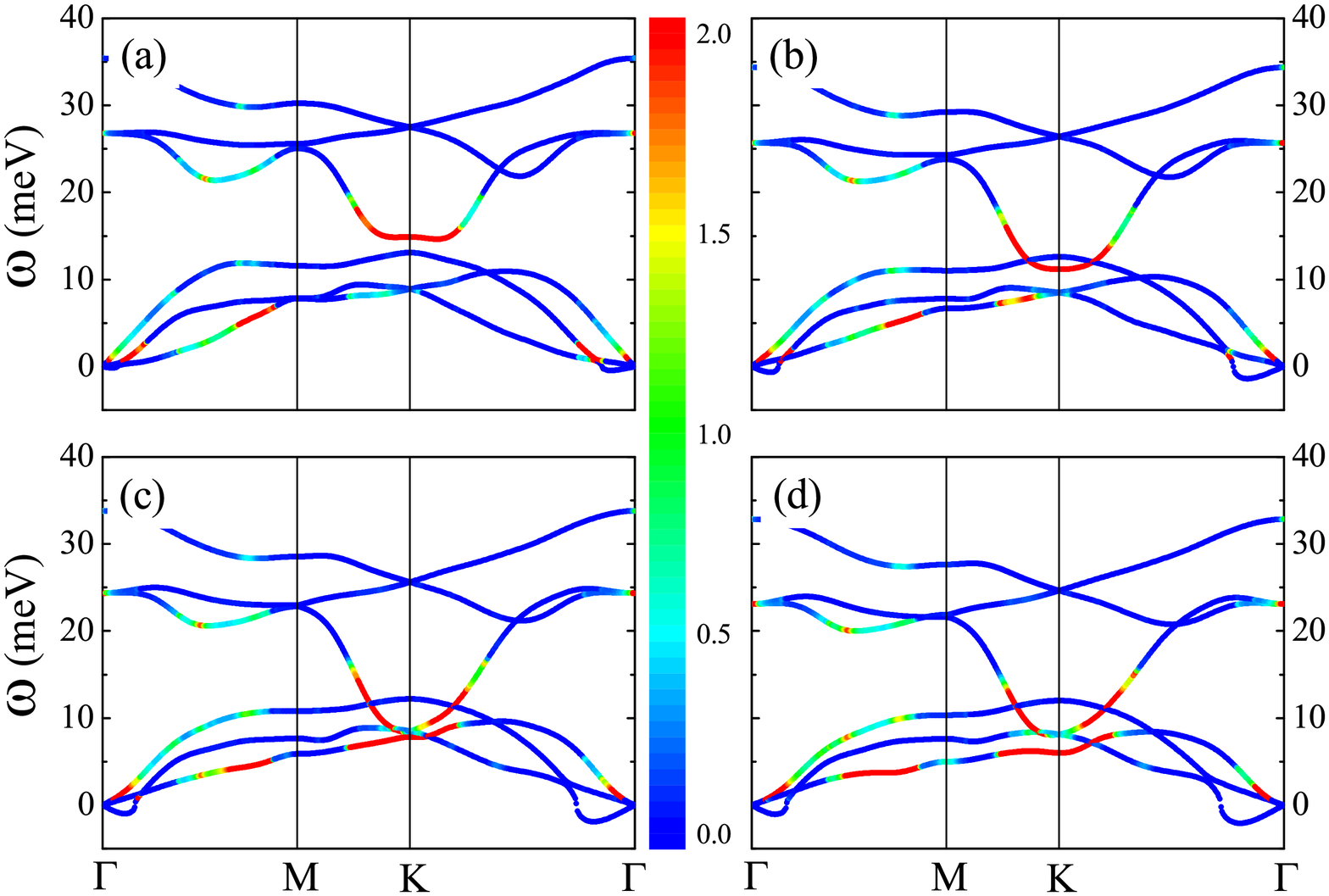}
\end{center}
  \caption{(Color online) Phonon spectra of arsenenes under (a) 2\%, (b) 4\%, (c) 6\%, and (d) 8\% BTS. The amplitude of $\lambda_{\mathbf{q \nu}}$ is mapped by different colors.}
\label{fig:phonon-s}
\end{figure}

Figure~\ref{fig:phonon-s} shows the phonon spectra of strained arsenenes. By applying the BTS, the phonon frequencies are softened, especially for the $A_1$ phonon mode
around the $K$ point.
Meanwhile, the EPC from the lowest acoustic phonon band between the $\Gamma$ and $M$ points becomes stronger and stronger with the increase of BTS.
Again, the phonon frequency in the lowest phonon band becomes imaginary around the $\Gamma$ point.
Further calculation, however, suggests that the 0.2 $e$/cell doped arsenene remains dynamically stable under 14\% BTS, slightly smaller than the critical strain, 18.4\%, as obtained in the undoped arsenene \cite{Zhang-Nanoscale7}.

\begin{figure}[b]
\begin{center}
\includegraphics[width=8.6cm]{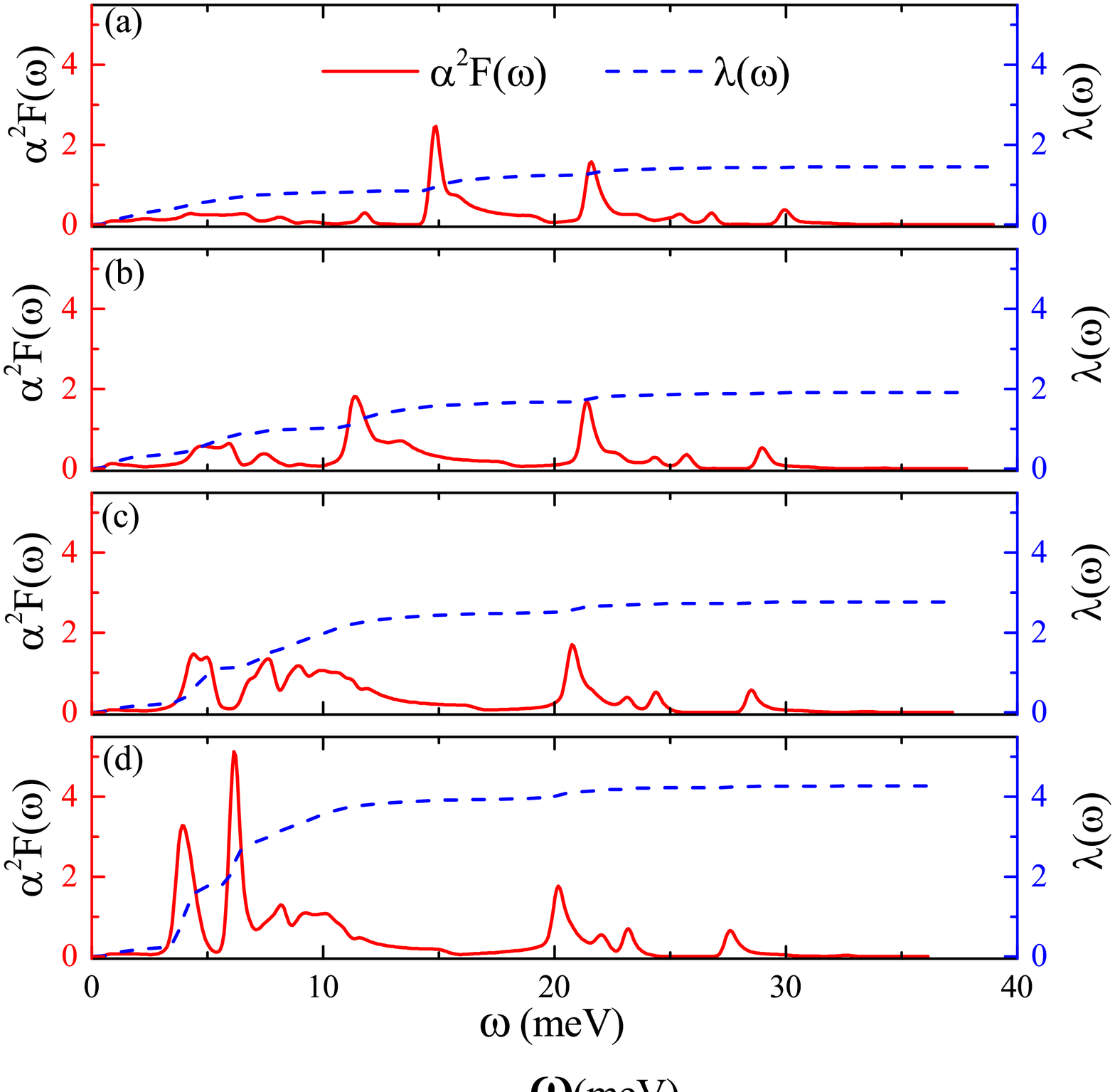}
\end{center}
  \caption{(Color online) Eliashberg spectral function $\alpha^2F(\omega)$ and $\lambda (\omega)$ the strained arsenenes with (a) 2\%, (b) 4\%, (c) 6\%, and (d) 8\% BTS.
  }
\label{fig:a2f-s}
\end{figure}

\begin{figure}[h]
\begin{center}
\includegraphics[width=8.6cm]{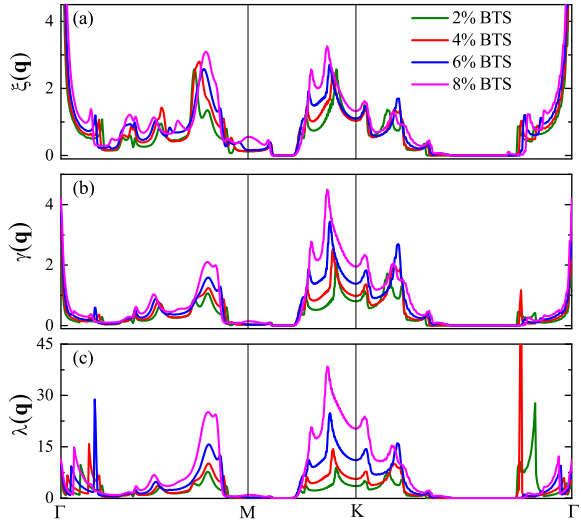}
\end{center}
  \caption{(Color online) $\xi(\mathbf{q})$,  $\gamma(\mathbf{q})$, and $\lambda(\mathbf{q})$ of strained arsenenes.}
\label{fig:nesting-s}
\end{figure}

When the applied BTS is less than 4\%, the two-peak structures of $\alpha^2F(\omega)$ are
preserved [Fig.~\ref{fig:a2f-s}(a) and Fig.~\ref{fig:a2f-s}(b)].
With the increase of BTS, $\alpha^2F(\omega)$ around 5 meV is enhanced.
This enhancement arises from the softening of the $A_1$ mode and the enhancement of the EPC in the lowest acoustic phonon band between $\Gamma$ and $M$.
We also calculate $\xi(\mathbf{q})$, $\gamma(\mathbf{q})$, and $\lambda(\mathbf{q}%
)$ for arsenene at different BTS [Fig.~\ref{fig:nesting-s}].
Similar to the case of the BTS-free arsenene, $\xi(\mathbf{q})$ is not a hegemonic factor
that determines $\lambda$.
In contrast, an obvious separation among the four curves in $\gamma(\mathbf{q})$ is observed. This separation results from the matrix element $|g_{\mathbf{k},\mathbf{q}\nu}^{ij}|$ around the Fermi level [Fig.~\ref%
{fig:nesting-s}(b)]. It is further enhanced by the softening of strongly coupled phonon modes [Fig.~\ref{fig:nesting-s}(c)], giving rise to a T$_c$ as high as 30.8 K at 8\% BTS.

To determine which electron band contributes most to the EPC, we calculate $\lambda_{\mathbf{k}i}$ defined by
\begin{equation}
\lambda_{\mathbf{k}i}=\frac{2}{\hbar N(0)N_q}\sum_{\mathbf{q}\nu j}\frac{1}{%
\omega_{\mathbf{q}\nu}}|g_{\mathbf{k},\mathbf{q}\nu}^{ij}|^2\delta(\epsilon_{%
\mathbf{k}}^i)\delta(\epsilon_{\mathbf{k+q}}^j),
\end{equation}
where $i$ is the band index of electrons.
$\lambda_{\mathbf{k}i}$ describes the scattering process of an electron from the $i$-th band to other bands by a phonon with momentum \textbf{q} and branch $\nu$.
It represents the contribution of electrons with momentum $\mathbf k$ at the $i$-th band to the EPC.

\begin{figure}[b]
\begin{center}
\includegraphics[width=8.6cm]{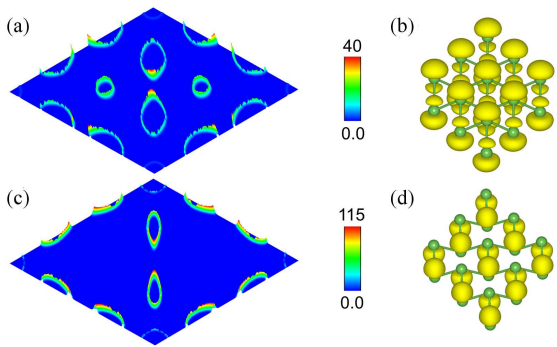}
\end{center}
  \caption{(Color online) (a) and (c): three-dimensional views of $\lambda_{\mathbf{k}i}$ in the reciprocal unit cell. (b) and (d): charge densities of electrons that couple most strongly with phonons. (a) and (b) correspond to 0.4 $e$/cell doping without BTS. (c) and (d) correspond to 0.2 $e$/cell doping with 8\% BTS.}
\label{fig:lambad-ki}
\end{figure}

For all the cases we have studied, we find that $\lambda_{\mathbf{k}i}$ behaves similarly.
Here we only show the results for the 0.4 $e$/cell doping without BTS [Fig.~\ref{fig:lambad-ki}(a)] and for the 0.2 $e$/cell doping with 8\% BTS [Fig.~\ref{fig:lambad-ki}(c)].
By comparing the contour picture of $\lambda_{\mathbf{k}i}$ with the Fermi surface shape shown in Fig.~\ref{fig:band-e}(h) and Fig.~\ref{fig:band-s}(h), we find that the main contribution to the EPC comes from the elliptical Fermi surface sheets.
In particular, the points near $M$ contribute most to the EPC.
The charge density of these electrons shows a $p_z$-like character [Fig.~\ref{fig:lambad-ki}(b) and Fig.~\ref{fig:lambad-ki}(d)], indicating that it is the $p_z$ orbital of arsenic atom that couples most strongly with phonons.

The maximal superconducting T$_c$ of 23.8 K is predicted for 4.65$\times$10$^{14}$cm$^{-2}$ electron-doped graphene under 16.5\% BTS.
Realistically, realizing such requirements in graphene may be very difficult in experiment. The advantage in the case of arsenene is that the high-T$_c$ superconductivity above 30 K may be obtained under a relatively smaller doping density (1.28$\times$10$^{14}$cm$^{-2}$) and BTS (12\%).
Recently, multilayer arsenenes were successfully grown on InAs using the
plasma-assisted process \cite{Tsai-Chem28}.
The bulk gray arsenic, which is the most stable phase among all arsenic allotropes \cite{Madelung}, could be used as a precursor to prepare arsenene \cite{Zhang-Angew54,Kamal-PRB91}.
By growing arsenene on a piezoelectric substrate, one can control BTS by applying a bias voltage to elongate or shorten the lattice constants \cite{Ding-Nano10}.
The electron doping can be achieved either by chemical doping or substitution, or by liquid or solid gating \cite{Yuan-AFM19,Fujimoto-PCCP15}.
Thus it is feasible to verify our prediction experimentally.

\section{CONCLUSION}

Based on the first-principles DFT electronic structure calculation, we predict that the semiconducting arsenene can become a phonon-mediated superconductor upon doping of electrons.
The maximal superconducting transition temperature is found to be around 10 K in the doped arsenene. It can be further lifted to 30 K by applying a 12\% BTS.
The superconducting pairing results mainly from the $A_1$ phonon mode around the $K$ point and the $p_z$-like electrons of arsenic atoms.

\begin{acknowledgments}
This work was supported by the National R\&D Program
of China (Grant No. 2017YFA0302901) and the
National Natural Science Foundation of China (Grants
Nos. 11474331, 11404383, and 11474004). M.G. is also
supported by the Zhejiang Provincial Natural Science Foundation (Grant No. LY17A040005) and by the K.C.Wong Magna Fund in Ningbo University.
\end{acknowledgments}


\begin{references}

\bibitem{Franceschi-Nature5}S. De Franceschi, L. Kouwenhoven, Ch. Sch\"{o}nenberger, and W. Wernsdorfer,
\textit{Hybrid superconductor-quantum dot devices},
Nat. Nanotech. {\bf 5}, 703 (2010).

\bibitem{Huefner-PRB79}M. Huefner, C. May, S. Ki\v{c}in, K. Ensslin, T. Ihn, M. Hilke, K. Suter, N. F. de Rooij, and U. Staufer,
\textit{Scanning gate microscopy measurements on a superconducting single-electron transistor},
Phys. Rev. B {\bf 79}, 134530 (2009).

\bibitem{Wang-CPL29}Q.-Y. Wang \textit{et al.},
\textit{Interface-Induced High-Temperature Superconductivity in Single Unit-Cell FeSe Films on SrTiO$_3$},
Chin. Phys. Lett. {\bf 29}, 037402 (2012).

\bibitem{Ge-Nat14}J.-F. Ge \textit{et al.},
\textit{Superconductivity above 100 K in single-layer FeSe films on doped SrTiO$_3$},
Nat. Mater. {\bf 14}, 285 (2015).

\bibitem{Hsu-PNAS105}F.-C. Hsu, J.-Y. Luo, K.-W. Yeh, T.-K. Chen, T.-W. Huang, P. M.
Wu, Y.-C. Lee, Y.-L. Huang, Y.-Y. Chu, D.-C. Yan, and M.-K.
Wu, \textit{Superconductivity in the PbO-type structure ¦Á-FeSe}, Proc.
Natl. Acad. Sci. USA {\bf 105}, 14262 (2008).

\bibitem{Lee-Nature515}J. J. Lee, F. T. Schmitt, R. G. Moore, S. Johnston, Y.-T. Cui, W. Li, M. Yi, Z. K. Liu, M. Hashimoto, Y. Zhang, D. H. Lu, T. P. Devereaux, D.-H. Lee, and Z.-X. Shen,
\textit{Interfacial mode coupling as the origin of the enhancement of T$_c$ in FeSe films on SrTiO$_3$},
Nature {\bf 515}, 245 (2014).

\bibitem{Uchoa-PRL98}B. Uchoa and A. H. Castro Neto,
\textit{Superconducting States of Pure and Doped Graphene},
Phys. Rev. Lett. {\bf 98}, 146801 (2007).

\bibitem{Profeta-Nature8}G. Profeta, M. Calandra, and F. Mauri,
\textit{Phonon-mediated superconductivity in graphene by lithium deposition},
Nat. Phys. {\bf 8}, 131 (2012).
%theoretical prediction graphene superconductivity.

\bibitem{Tiwari-arXiv}A. P. Tiwari, S. Shin, E. Hwang, S.-G. Jung, T. Park, and H. Lee,
\textit{Superconductivity at 7.4 K in Few Layer Graphene by Li intercalation},
arXiv:1508.06360.
%7.4 K, Li intercalated graphene.

\bibitem{Chapman-arXiv}J. Chapman, Y. Su, C. A. Howard, D. Kundys, A. Grigorenko, F. Guinea, A. K. Geim, I. V. Grigorieva, and R. R. Nair,
\textit{Superconductivity in Ca-doped graphene},
Sci. Rep. {\bf 6}, 23254 (2016).
% 6K Ca-doped graphene

\bibitem{Si-PRL111}C. Si, Z. Liu, W. Duan, and F. Liu,
\textit{First-Principles Calculations on the Effect of Doping and Biaxial Tensile Strain on Electron-Phonon Coupling in Graphene},
Phys. Rev. Lett. {\bf 111}, 196802 (2013).

\bibitem{Liu-Nano8}H. Liu, A.T. Neal, Z. Zhu, Z. Luo, X. Xu, D. Tom\'{a}nek, and P. D. Ye,
\textit{Phosphorene: An Unexplored 2D Semiconductor with a High Hole Mobility},
Nano Lett. {\bf 8}, 4033 (2014).

\bibitem{Lalmi-APL97}B. Lalmi, H. Oughaddou, H. Enriquez, A. Kara, S. Vizzini, B. Ealet, and B. Aufray,
\textit{Epitaxial growth of a silicene sheet},
Appl. Phys. Lett. {\bf 97}, 223109 (2010).

\bibitem{Chen-PRL109}L. Chen, C.-C. Liu, B. Feng, X. He, P. Cheng, Z. Ding, S. Meng, Y. Yao, and K. Wu,
\textit{Evidence for Dirac Fermions in a Honeycomb Lattice Based on Silicon},
Phys. Rev. Lett. {\bf 109}, 056804 (2012).

\bibitem{Mannix-Science}A. J. Mannix \textit{et al.,}
\textit{Synthesis of borophenes: Anisotropic, two-dimensional boron polymorphs},
Science {\bf 350}, 1513 (2015).

\bibitem{Feng-arXiv1}B. Feng \textit{et al.},
\textit{Experimental Realization of Two-Dimensional Boron Sheets},
Nat. Chem. {\bf 8}, 563 (2016).

\bibitem{Chen-PRL110}L. Chen, H. Li, B. Feng, Z. Ding, J. Qiu, P. Cheng, K. Wu, and S. Meng,
\textit{Spontaneous Symmetry Breaking and Dynamic Phase Transition in Monolayer Silicene},
Phys. Rev. Lett. {\bf 110}, 085504 (2013).

\bibitem{Nie-APL94}Y. F. Nie, E. Brahimi, J. I. Budnick, W. A. Hines, M. Jain, and B. O. Wells,
\textit{Suppression of superconductivity in FeSe films under tensile strain},
Appl. Phys. Lett. {\bf 94}, 242505 (2009).

\bibitem{Ding-Nano10}F. Ding, H. Ji, Y. Chen, A. Herklotz, K. D\"{o}rr, Y. Mei, A. Rastelli, and O. G. Schmidt,
\textit{Stretchable Graphene: A Close Look at Fundamental Parameters through Biaxial Straining},
Nano Lett. {\bf 10}, 3453 (2010).

\bibitem{Conley-Nano13}H. J. Conley, B. Wang, J. I. Ziegler, R. F. Haglund Jr., S. T. Pantelides, and K. I. Bolotin,
\textit{Bandgap Engineering of Strained Monolayer and Bilayer MoS$_2$},
Nano Lett. {\bf 13}, 3626 (2013).

\bibitem{Wan-EPL104}W. Wan, Y. Ge, F. Yang, and Y. Yao,
\textit{Phonon-mediated superconductivity in silicene predicted by first-principles density functional calculations},
Europhys. Lett. {\bf 104}, 36001 (2013).

\bibitem{Shao-EPL108}D. F. Shao, W. J. Lu, H. Y. Lv, and Y. P. Sun,
\textit{Electron-doped phosphorene: A potential monolayer superconductor},
Europhys. Lett. {\bf 108}, 67004 (2014).

\bibitem{Ge-NJP17}Y. Ge, W. Wan, F. Yang, and Y. Yao,
\textit{The strain effect on superconductivity in phosphorene: a first-principles prediction},
New J. Phys. {\bf 17}, 035008 (2015).

\bibitem{Penev-Nano16}E. S. Penev, A. Kutana, and B. I. Yakobson, \textit{Can twodimensional boron superconduct?}, Nano Lett. {\bf 16}, 2522 (2016).

\bibitem{Gao-PRB95}M. Gao, Q.-Z. Li, X.-W. Yan, and J. Wang,
\textit{Prediction of phonon-mediated superconductivity in borophene},
Phys. Rev. B {\bf 95}, 024505 (2017).

\bibitem{Cheng-2D4}C. Cheng, J.-T. Sun, H. Liu, H.-X. Fu, J. Zhang, X.-R. Chen, and S. Meng,
\textit{Suppressed superconductivity in substrate-supported $\beta_{12}$ borophene by tensile strain and electron doping},
2D Mater. {\bf 4}, 025032 (2017).

\bibitem{Zhang-Angew54}S. Zhang, Z. Yan, Y. Li, Z. Chen, and H. Zeng,
\textit{Atomically Thin Arsenene and Antimonene: Semimetal-Semiconductor and Indirect-Direct Band-Gap Transitions},
Angew. Chem. Int. Ed. {\bf 54}, 3112 (2015).

\bibitem{Kamal-PRB91}C. Kamal and M. Ezawa,
\textit{Arsenene: Two-dimensional buckled and puckered honeycomb arsenic systems},
Phys. Rev. B {\bf 91}, 085423 (2015).

\bibitem{Pizzi-Nat7}G. Pizzi, M. Gibertini, E. Dib, N. Marzari, G. Iannaccone, and G. Fiori,
\textit{Performance of arsenene and antimonene double-gate MOSFETs from first principles},
Nat. Communs. {\bf 7}, 12585 (2016).

\bibitem{Zhang-Nanoscale7}H. Zhang, Y. Ma, and Z. Chen,
\textit{Quantum spin hall insulators in strain-modified arsenene},
Nanoscale {\bf 7}, 19152 (2015).

\bibitem{pwscf}P. Giannozzi {\it et al.},
\textit{QUANTUM ESPRESSO: a modular and open-source software project for quantum simulations of materials},
J. Phys.: Condens. Matter {\bf 21}, 395502 (2009).
%http://www.quantum-espresso.org.

\bibitem{Troullier-PRB43}N. Troullier and J. L. Martins,
\textit{Efficient pseudopotentials for plane-wave calculations}, Phys. Rev. B {\bf 43}, 1993 (1991).

\bibitem{Giustino-PRB76}F. Giustino, M. L. Cohen, and S. G. Louie,
\textit{Electron-phonon interaction using Wannier functions},
Phys. Rev. B {\bf 76}, 165108 (2007).

\bibitem{Methfessel-PRB40}M. Methfessel and A. T. Paxton,
\textit{High-precision sampling for Brillouin-zone integration in metals},
Phys. Rev. B {\bf 40}, 3616 (1989).

\bibitem{Baroni-RMP73_515}S. Baroni, S. de Gironcoli, A. Dal Corso, and P. Giannozzi,
\textit{Phonons and related crystal properties from density-functional perturbation theory},
Rev. Mod. Phys. {\bf 73}, 515 (2001). %DFPT

\bibitem{Marzari-PRB56}N. Marzari and D. Vanderbilt,
\textit{Maximally localized generalized Wannier functions for composite energy bands},
Phys. Rev. B {\bf 56}, 12847 (1997).

\bibitem{Souza-PRB65}I. Souza, N. Marzari, and D. Vanderbilt,
\textit{Maximally localized Wannier functions for entangled energy bands},
Phys. Rev. B {\bf 65}, 035109 (2001).

\bibitem{Mostofi-CPC178}A. A. Mostofi, J. R. Yates, Y.-S. Lee, I. Souza, D. Vanderbilt, and N. Marzari,
\textit{wannier90: A tool for obtaining maximally-localised Wannier functions},
Comput. Phys. Commun. {\bf 178}, 685 (2008).
A. A. Mostofi, J. R. Yates, G. Pizzi, Y.-S. Lee, I. Souza, D. Vanderbilt, and N. Marzari,
\textit{An updated version of wannier90: A tool for obtaining maximally-localised Wannier functions},
Comput. Phys. Commun. {\bf 185}, 2309 (2014).

\bibitem{Noffsinger-CPC181}J. Noffsinger, F. Giustino, B. D. Malone, C.-H. Park, S. G. Louie, and M. L. Cohen,
\textit{EPW: A program for calculating the electron¨Cphonon coupling using maximally localized Wannier functions},
Comput. Phys. Commun. {\bf 181}, 2140 (2010).
S. Ponc\'{e}, E. R. Margine, C. Verdi, and F. Giustino,
\textit{EPW: Electron-phonon coupling, transport and superconducting properties using maximally localized Wannier functions},
Comput. Phys. Commun. {\bf 209}, 116 (2016).

\bibitem{Allen-PRB6_2577}P. B. Allen,
\textit{Neutron spectroscopy of superconductors},
Phys. Rev. B {\bf 6}, 2577 (1972).

\bibitem{Allen-RPB12_905}P. B. Allen and R. C. Dynes,
\textit{Transition temperature of strong-coupled superconductors reanalyzed},
Phys. Rev. B {\bf 12}, 905 (1975).

\bibitem{Cahangirov-PRL102}S. Cahangirov, M. Topsakal, E. Akt\"{u}rk, H. \c{S}ahin, and S. Ciraci,
\textit{Two- and One-Dimensional Honeycomb Structures of Silicon and Germanium},
Phys. Rev. Lett. {\bf 102}, 236804 (2009).

\bibitem{Sahin-PRB80}H. \c{S}ahin, S. Cahangirov, M. Topsakal, E. Bekaroglu, E. Akturk, R. T. Senger, and S. Ciraci,
\textit{Monolayer honeycomb structures of group-IV elements and III-V binary compounds:
First-principles calculations},
Phys. Rev. B {\bf 80}, 155453 (2009).

\bibitem{Tsai-Chem28}H.-S. Tsai, S.-W. Wang, C.-H. Hsiao, C.-W. Chen, H. Ouyang, Y.-L. Chueh, H.-C. Kuo, and J.-H. Liang,
\textit{Direct Synthesis and Practical Bandgap Estimation of Multilayer Arsenene Nanoribbons},
Chem. Mater. {\bf 28}, 425 (2016).

\bibitem{Madelung} O. Madelung, \textit{Semiconductors: Data Handbook}, 3rd ed.; Springer-Verlag: New York, pp 405-411 (2004).

\bibitem{Yuan-AFM19}H. Yuan, H. Shimotani, A. Tsukazaki, A. Ohtomo,
M. Kawasaki, and Y. Iwasa,
\textit{High-density carrier accumulation in ZnO field-effect transistors gated by electric double layers of ionic liquids},
Adv. Funct. Mater. {\bf 19}, 1046 (2009).

\bibitem{Fujimoto-PCCP15}T. Fujimoto and K. Awaga,
\textit{Electric-double-layer field-effect transistors with ionic liquids},
Phys. Chem. Chem. Phys. {\bf 15}, 8983 (2013).


\end{references}
\end{document}